\documentclass [10pt,twoside]{article}

\usepackage{cite}
\usepackage{graphicx}
\usepackage[usenames]{color}
\usepackage{subfigure}
\usepackage{setspace}
\usepackage{amssymb}
\usepackage{multirow}

\usepackage[hang,small]{caption}

\usepackage{amsmath}                  

\usepackage{geometry}
\usepackage{fancyhdr}
\newcommand{\be}{\begin{equation}}
\newcommand{\ee}{\end{equation}}
\newcommand{\ba}{\begin{eqnarray}}
\newcommand{\ea}{\end{eqnarray}}
\newcommand{\bc}{\begin{center}}
\newcommand{\ec}{\end{center}}
\usepackage{titlesec,titletoc}
  \titleformat{\section}{\Large\sf\bfseries}{\thesection}{1em}{}
  \titleformat{\subsection}{\large\sf\bfseries}{\thesubsection}{1em}{}

\title{\sf\bfseries \ntitle}
\author{\normalsize Inyong Cho \footnote{email: iycho@seoultech.ac.kr}, \  Naveen K. Singh\footnote{email: naveen.nkumars@gmail.com}}
\date{}

\newcommand{\pghdr}{\footnotesize {Inyong Cho and Naveen K. Singh}  -- Unimodular Theory of Gravity and Inflation }
\newcommand{\ntitle}{Unimodular Theory of Gravity and Inflation}

\begin{document}

\vspace{-3cm}
\maketitle
\vspace{-0.6cm}
\bc
\small{ Institute of Convergence Fundamental Studies \& School of Liberal Arts, \\ Seoul National University of Science and Technology, 
Seoul 139-743, Korea}
\ec

\bc\begin{minipage}{0.9\textwidth}\begin{spacing}{1}{\small {\bf Abstract:}
  We investigate inflation  and its scalar perturbation driven by a massive scalar field in the unimodular theory of gravity. We introduce a parameter $\xi$ with which
 the theory is invariant under general unimodular coordinate transformations.
When the unimodular parameter is $\xi=6$,
the classical picture of inflation is reproduced in the unimodular theory
because it recovers the background equations
of the standard theory of general relativity.  We show that for $\xi=6$, the theory is  equivalent to the standard theory of general relativity
at the perturbation level. Unimodular gravity constrains the gauge degree of freedom in the scalar perturbation,
but the perturbation equations are similar to  those in general relativity.
For $\xi \neq 6$, we derive the power spectrum and the spectral index,
and obtain the unimodular correction to the tensor-to-scalar ratio. Depending on the value of $\xi$, the correction can either raise or lower
the value of the tensor-to-scalar ratio. 


}\end{spacing}\end{minipage}\ec

\section{Introduction}
The de Sitter expansion  in the period of inflation, provides a solution to the flatness problem, the horizon problem, 
the entropy problem etc  \cite{guth,starobinsky1,starobinsky2,kazanas,sato,linde1,linde2,linde3,albrecht}.  The inflationary model also
explains the structure formation of the Universe by considering the perturbations which are generated in the period of
inflation.
Not only for the period of inflation in the early universe, the role of the exponential expansion of the Universe is important in the current era of
the accelerating universe \cite{Perlmutter,Riess}. There are many theories which describe the current accelerating universe such as DGP model, $f(R)$-gravity and scalar 
field models, etc. The scalar field model
is one of the simplest model. The cosmological constant is also one of the explanation for the current acceleration of the Universe. However, 
 adding the cosmological constant term, the theory suffers from the fine tuning problem.
 
 The unimodular theory of gravity was initially developed in Refs. \cite{Einstein, Anderson}. One of the motivations of considering the unimodular theory of gravity is
 to solve the cosmological constant problem \cite{Weinberg}. Another interesting implication is that it explains the current expansion of the Universe by
 considering only single component such as the cosmological constant, or by the non-relativistic matter \cite{Jain:2011,Jain:2012gc}. 
 The full metric is decomposed in the unimodular metric and a scalar field \cite{Jain:2011,Jain:2012gc}.  The value of the determinant of the unimodular metric is the same as the
 determinant of the Minkowski metric. The basic idea of the unimodular theory of gravity is to consider the determinant
 of metric not as  a dynamical variable \cite{Ng1,Zee,Buchmuller,Weinberg}, and hence the  cosmological constant term is absent from the action. However,
 it was shown in  Ref. \cite{Weinberg} that the cosmological constant appears as an integration constant in this theory. Some progress in the unimodular
 theory of gravity has been discussed in Refs. \cite{Henneaux89,Unruh1989,Ng2,Finkelstein,Tiwari:2003yq,Alvarez05,Alvarez06,Abassi,Ellis,Jain:2012cw,Singh:2012sx,Gao:2014nia,Kluson:2014esa,
 Padilla:2014yea,Saltas:2014cta,Barcelo:2014mua,Barcelo:2014qva}.   In the case of the unimodular theory we have the unimodular constraint equation, and hence
 it reduces the gauge degree of freedom. Therefore, the perturbations are determined  by the standard equation of motion with the constraint equation. In GR, we seek for the gauge 
 invariant scalar perturbation, i.e., the curvature perturbation.  This quantity is still an invariant quantity in unimodular gravity since gauge transformations are not related 
 with the determinant of metric.
 
 In the recent paper \cite{Gao:2014nia}, the authors developed the theory of cosmological perturbations considering the  unimodular constraint, i.e., the fixed determinant of the metric.
 They considered the variation of the metric determinant is zero up to the linear order in the perturbation, and discussed the cosmological perturbations in the
 matter and the radiation dominated eras.  We adopt their method in this paper up to the linear order and  get the power spectrum  of the  cosmological  
 perturbation produced during inflation.

 In general, all the metric components are dynamical fields. However, we can assume that some of them are not dynamical \cite{Weinberg}. In the unimdular theory,
we have the determinant of the metric fixed, i.e., $g_{\mu\nu}\delta g^{\mu\nu}=0$. All the components are  dynamical and adjusted 
such that they satisfy $\sqrt{-g}$ is fixed. Without any constraint, the gravitational action with the cosmological constant
term provides us the standard Einstein equation,
\ba
R_{\mu\nu}-\frac{1}{2}g_{\mu\nu}R + \Lambda g_{\mu\nu}=  8 \pi G T_{\mu\nu},
\ea
where $R_{\mu\nu}$, $R$ and $T_{\mu\nu}$ are the Ricci tensor, the Ricci scalar, and the energy-momentum tensor of matter. Applying the unimodular constraint, in unimodular gravity we have to
subtract all the terms which are proportional to metric $g_{\mu\nu}$ from
the Einstein equation. Therefore, we have
\ba
R_{\mu\nu} - \frac{1}{4}g_{\mu\nu} R = 8 \pi G \left(T_{\mu\nu} - \frac{1}{4} g_{\mu\nu}T\right).
\ea
This is the standard field equation in unimodular gravity. In this paper, we discuss  unimodular gravity in a slightly different approach \cite{Jain:2011,Jain:2012gc}.
We first decompose the full metric into the unimodular metric and a scalar field, and write the whole action in terms of these fields  with introducing an 
additional parameter $\xi$. 
 
 The paper is organized as follows. In  Sec. 2, we describe  the proposed model of the unimodular theory of gravity \cite{Jain:2011,Jain:2012gc} in brief. In  Sec. 3, we derive the background
 field equations to show that inflation can be reproduced in the unimodular theory in the same way as in GR. In Sec. 4, we discuss the cosmological perturbations. In Sec. 5, we solve  
 the perturbation equation and calculate the power spectrum 
 and the spectral index. In Sec. 6, we conclude.

\section{Unimodular Theory of Gravity} \label{uni_intro}
   We decompose the full metric $g_{\mu\nu}$ into two parts \cite{Jain:2011,Jain:2012gc} as
\begin{eqnarray}
g_{\mu\nu}=\mathcal{A}^2 \bar{g}_{\mu\nu} \ \ \ \mbox{and} \ \ \ g^{\mu\nu}=\frac{1}{\mathcal{A}^2} \bar{g}^{\mu\nu},
\label{full_g}
\end{eqnarray}
where $\bar{g}_{\mu\nu}$ and $\bar{g}^{\mu\nu}$ are the metric and the inverse metric corresponding to  unimodular gravity. $\mathcal{A}$ is
a scalar field which will turn out to be the scale factor  of the Universe. The unimodular metric satisfies $\sqrt{-\bar{g}_{\mu\nu}}= f(x)$, where
$f(x)$ is the determinant of the Minkowski metric. Using Eq. (\ref{full_g}), we can decompose the Christoffel symbol and the Ricci tensor as follows,

\ba
\Gamma^\mu_{\alpha\beta} = \bar \Gamma^\mu_{\alpha\beta} +\tilde
\Gamma^\mu_{\alpha\beta} \ , \ \  R_{\mu\nu} = \bar R_{\mu\nu} + \tilde R_{\mu\nu} ,
\label{christ_Rmunu}
\ea
where $\bar \Gamma^\mu_{\alpha\beta}$ and $\bar R_{\mu\nu}$  are computed with the metric $\bar g_{\mu\nu}$ and
\ba
\tilde\Gamma^\mu_{\alpha\beta} &=& \bar g^\mu_\beta\, \partial_\alpha \ln\mathcal{A}
 + \bar g^\mu_\alpha\, \partial_\beta \ln\mathcal{A} - \bar g_{\alpha\beta}\,
\partial^\mu \ln\mathcal{A} ,\label{eq:tilde_Gamma} \\
\tilde R_{\mu\nu} &=& -\tilde \Gamma^\alpha_{\mu\nu;\alpha}
+ \tilde \Gamma^\alpha_{\mu\alpha;\nu} - \tilde \Gamma^\alpha_{\beta\alpha}
\tilde \Gamma^\beta_{\mu\nu} + \tilde \Gamma^\alpha_{\beta\nu}
\tilde \Gamma^\beta_{\mu\alpha} \nonumber \\
&=& 2(\ln\mathcal{A})_{;\mu\nu} + \bar g_{\mu\nu}
(\ln\mathcal{A})^{ \ \ \alpha}_{;\alpha} - 2\partial_\mu\ln\mathcal{A} \,
\partial_\nu\ln\mathcal{A} + 2 \bar g_{\mu\nu} \partial_\beta\ln\mathcal{A}\, 
\partial^\beta\ln\mathcal{A} .
\ea
Using the above definitions, one can also compute the Ricci scalar as
\be
R = g^{\mu\nu} R_{\mu\nu} = \frac1{\mathcal{A}^2}(\bar R + \tilde R),
\ee 
where
\be
\tilde R = 6 (\ln\mathcal{A})^{ \ \ \mu}_{;\mu} + 6\partial_\mu \ln\mathcal{A}
\ \partial^\mu\ln\mathcal{A} .
\ee 
Then the gravitational action can be written as
\be
S_{\rm E} = \int d^4x \sqrt{-\bar g} {1\over 16 \pi G}
\, \left[\mathcal{A}^2 \bar R + \mathcal{A}^2 \tilde R\right].
\ee
In the terms of the scalar field, this can be written as
\be
S_{\rm E} = \int d^4x \sqrt{-\bar g} \frac{1}{16 \pi G}
\, \left[\mathcal{A}^2 \bar R - 6 \partial_\alpha\mathcal{A} \partial^\alpha\mathcal{A}\right].
\label{eq_SE}
\ee
 By replacing the coefficient 6 with  a new parameter $\xi$ in the action  (\ref{eq_SE}), we consider a new action as follows,
\be
S_{\rm uni} = \int d^4x \sqrt{-\bar g} \frac{1}{16 \pi G}
\, \left[\mathcal{A}^2 \bar R - \xi \partial_\alpha\mathcal{A} \partial^\alpha\mathcal{A}\right].
\label{eq_SE1}
\ee
For $\xi=6$, the theory is equivalent to the standard GR. For $\xi\neq 6$, the action is  invariant  under the general unimodular  coordinate transformations, and $\mathcal{A}$ is
treated as a scalar field \cite{Jain:2011,Jain:2012gc}. Varying the above action with respect to the metric $\bar{g}^{\mu\nu}$ and the field $\mathcal{A}$, we get the corresponding
equations of motion as
\ba
\mathcal{A}^2 \left[\bar R_{\mu\nu} - {1\over 4} \bar g_{\mu\nu}\bar R\right]
&+& \left[\left(\mathcal{A}^2\right)_{;\mu\nu} - {1\over 4}\bar g_{\mu\nu} 
\left(\mathcal{A}^2\right)^{ \ \ \lambda}_{;\lambda} \right]
- \xi \left[\partial_\mu\mathcal{A}\partial_\nu\mathcal{A} -{1\over 4}\bar g_{\mu\nu}
\partial^\lambda\mathcal{A}\partial_\lambda\mathcal{A} \right]\nonumber\\
 &=& -{8 \pi G} \left[T_{\mu\nu}
- {1\over 4} \bar g_{\mu\nu} T\right] ,
\label{eq_Rmunu}
\ea
and 
\be
2 \mathcal{A}\bar R + 2 \xi \bar g^{\mu\nu} \mathcal{A}_{;\mu\nu} = \kappa T_\mathcal{A}, \label{eq_A}
\ee
where $\kappa= 16 \pi G$, $T_{\mu\nu}$ is the energy-momentum tensor of the matter field, $T=T^{\lambda}_{\lambda}$, and $T_\mathcal{A}$ includes all the contributions from the coupling of
$\mathcal{A}$ with matter fields.  In Secs. {\ref{inflation}}, {\ref{Pert}} and {\ref{power}}, we shall derive the physical quantities at the background level as well
as at the perturbation level in terms of $\xi$.
\section{Inflation} \label{inflation}
The equations (\ref{eq_Rmunu}) and (\ref{eq_A}) have been used to describe the expansion of the Universe for the current era in Refs. \cite{Jain:2011,Jain:2012gc} in which
authors discussed that the expansion of the Universe can be described  only by the cosmological constant, or the nonrelativistic matter. In contrast to
the standard $\Lambda$CDM model, it sounds interesting since only one component of the energy density can explain the dynamics of the Universe.
In this paper,  we consider a scalar field  $\phi$ which is responsible for inflation. The matter-field action  is given by
\be
S = \int d^4x \sqrt{-\bar g} \left[{1\over 2}\mathcal{A}^2\bar g^{\mu\nu} \partial_\mu\phi
\partial_\nu\phi + {1\over 2} m^2\mathcal{A}^4\phi^2\right].
\label{matter_action}
\ee
Here, we consider the background unimodular metric as $\bar{g}_{\mu\nu}= \mbox{diag}(-1,1,1,1)$. For the above action (\ref{matter_action}), the energy-momentum tensor
of the matter field is given by
\ba
T_{\mu\nu} &=& \mathcal{A}^2 \partial_\mu \phi\partial_\nu\phi,
\ea
and 
$T_{\mathcal{A}}$ is defined as
\ba
T_\mathcal{A} &\!=\!& -\mathcal{A} \bar g^{\mu\nu} \partial_\mu\phi\partial_\nu\phi
- 2\mathcal{A}^3 m^2\phi^2.
\ea
For $\xi=6$, the $(0,0)$ component of Eq. (\ref{eq_Rmunu}) gives 
\ba
\frac{\mathcal{A}''}{\mathcal{A}} - 2 \left(\frac{\mathcal{A'}}{\mathcal{A}}\right)^2 = - 4 \pi G \phi'^2, \label{slow1}
\ea
where the prime denotes the derivative with respect to the conformal time $\eta$. The equation (\ref{eq_A}) gives
\ba
\frac{\mathcal{A}''}{\mathcal{A}} = \frac{8 \pi G}{3} \mathcal{A}^2 \Big[- \frac{\phi'^2}{2 \mathcal{A}^2} + 2 V(\phi)\Big], \label{eq2_A}
\ea
where $V(\phi)= m^2 \phi^2/2$. Now let us consider $\mathcal{A}=a(t)$, so from Eq. (\ref{slow1}) we can write the derivative of the Hubble parameter $H=\dot{a}/a$ as
\ba
\dot{H} = - 4 \pi G \dot{\phi}^2, \label{FR1}
\ea
where the overdot denotes the derivative with respect to the cosmological time $t$ defined by the transformation $dt = a d\eta$. From Eqs. (\ref{slow1}) and (\ref{eq2_A}), we also recover the first Friedman equation
\ba
H^2 = \frac{8 \pi G}{3} \left(\frac{\dot{\phi}^2}{2 } + V\right) \ . \label{FR2}
\ea
The slow-roll parameter $\epsilon$ is given by, as usual,
\ba
\epsilon = - \frac{\dot{H}}{H^2} = 4 \pi G \frac{\dot{\phi}^2}{H^2}.
\ea
The Friedmann equations (\ref{FR1}) and (\ref{FR2}) obtained in unimodular gravity are  in the standard form as in GR. The slow-roll parameters are defined 
also in the same way as in  GR. If the kinetic term of the scalar field is much smaller than the potential term, the scale factor exhibits an exponential expansion. The slow-roll parameters
determine the sufficient period for inflation which solves the  problems of the standard Big Bang cosmology. 

Now let us consider the general case of $\xi$. In this case, from Eq. (\ref{eq_Rmunu}) we have
\ba
\frac{a''}{a}- \frac{\xi-2}{2}\left(\frac{a'}{a}\right)^2 = - 4 \pi G \phi'^2 \label{eq3_A}.
\ea
If we introduce $\delta \equiv \xi- 6$, with $\delta$ of an order of unity,  we have
\ba
\dot{H}= \frac{\delta}{2} H^2 - 4 \pi G \dot{\phi}^2 .
\ea
 The equation (\ref{eq_A}) gives
\ba
\xi \frac{a''}{a} =  \kappa a^3 \Big[- \frac{\phi'^2}{2 a^2} + 2 V(\phi)\Big] \label{eq4_A}.
\ea
Using Eqs. (\ref{eq3_A}) and  (\ref{eq4_A}), we have
\ba
H^2 &=& \frac{64 \pi G}{\xi (\xi-2)} \Big[\frac{(\xi-2)}{4}\frac{\dot{\phi}^2}{2} + V\Big] 
    \approx \frac{8 \pi G}{3} \Big[\left(1+ \frac{\delta}{4}\right)\frac{\dot{\phi}^2}{2}+ V \Big] \left(1+ \frac{ \delta}{6}\right)^{-1}\left(1+ \frac{ \delta}{4}\right)^{-1} \nonumber \\
    &\approx& \left(1+ \frac{ \delta}{6}\right)^{-1}\left(1+ \frac{ \delta}{4}\right)^{-1}  H_{\mbox{\tiny GR}}^2,  \label{HUB}
  \ea
where the last approximation is for the slow-roll region, and $H_{\mbox{\tiny GR}} \approx \sqrt{(8 \pi G) V/3}$ is the Hubble parameter in  GR. The form of equation of motion for $\phi$ is  as usual since the parameter $\xi$ is only used in the gravitational action.
The slow-roll parameter becomes
\ba
\epsilon = -\frac{\dot{H}}{H^2}= \epsilon_1 - \frac{\delta}{2}, \label{mainslow}
\ea
where\ba
\epsilon_1 &=& 4 \pi G \frac{\dot{\phi}^2}{H^2} \approx \frac{1}{16 \pi G} \left(\frac{V'}{V}\right)^2 \left(1+ \frac{ \delta}{6}\right)^{2}\left(1+ \frac{ \delta}{4}\right)^{2}
= \frac{2 M_p^2}{\phi^2} \left(1+ \frac{ \delta}{6}\right)^{2}\left(1+ \frac{ \delta}{4}\right)^{2}  \ .
\ea
 For $\xi=6 \ (\delta=0)$, the slow-roll parameter $\epsilon_1  = 2 M_p^2/\phi^2$  is the same as in the standard GR. The parameter $\delta$ modifies the slow-roll 
parameter $\epsilon$ via Eq. (\ref{mainslow}), and thus the number of $e$-foldings. The Hubble parameter is also slightly modified. The 
negative small value of $\delta$ increases the value of Hubble parameter. Using Eq. (\ref{mainslow}), one can write,
\ba
\frac{\ddot{a}}{a} = \left(1+ \frac{\delta}{2}- \epsilon_1\right) H^2 \label{acce}.
\ea
From this, the necessary condition for inflation $\ddot{a}>0$ puts a lower bound on the parameter, $\delta > -2$ ($\xi>4$). 
\section{Cosmological Perturbations} \label{Pert}
In  this section, we discuss the cosmological perturbations. The decomposition of the perturbations into the scalar, the vector and the tensor fields is still valid in the unimodular theory. We 
consider only the scalar perturbations in this paper. The perturbed metric and scalar field are defined as 
\ba 
 ds_g^2 &=& a^2\Big\{(-1- 2 A) d\eta^2 +  2 B_{,i}dx^i d\eta + \Big[(1-2\psi)\delta_{il} +  D_{il}E \Big] dx^i d x^l \Big\} \equiv a^2 ds_{\bar{g}}^2, \\
 \phi &=& \chi_0(\eta) + \chi_1(\eta, \vec{x}),
\ea
where $A, \psi, E$ and  $B$ are scalar perturbation fields, $D_{ij}\equiv \partial_i\partial_j - (1/3) \delta_{ij}\bigtriangledown^2$, and $ds_g$ is the physical line element. We perturb each term in Eq. (\ref{eq_Rmunu}) in the linear order, and  write all the components,
$(0,0)$, $(i,0)$ and $(i,j)$ of these terms.  In the unimodular theory $\sqrt{-\bar{g}}$ is not dynamical, but the line element which is governed by the full metric $g_{\mu\nu}$,
is still invariant under general coordinate transformations. Therefore, we have the gauge dependent gravitational potential, $\psi$, $A$, etc., same as in GR. In the linear order, the
constraint $\delta \sqrt{-g}=0$ gives
\ba
A - 3 \psi + E_{,ii}=0 \label{constraint1}.
\ea
For the background, $\bar{R}_{\mu\nu}$ and $\bar{R} =0$, since the background metric $\bar{g}_{\mu\nu}$ is the Minkowski metric. Only the perturbed part of these can 
survive. The $(0,0)$, $(i,j)$ and $(0,i)$ components of the perturbed part of $\mathcal{A}^2 \left( \bar{R}_{\mu\nu}- \frac{1}{4}\bar{g}_{\mu\nu}\bar{R}\right) $ are given by
\ba
\mathcal{A}^2\left(\bar{R}_{00}- \frac{1}{4}\bar{g}_{00}\bar{R}\right)&=& \frac{\mathcal{A}^2}{2} \Big[  (A+ B'+ 2 \psi)^{, i}_{ \ \  i}
+ 3 \psi '' + \frac{1}{2} ( D^k_i E)^{,i}_{ \ \  k}\Big], \label{00part1}\\ 
\mathcal{A}^2\left(\bar{R}_{ij}- \frac{1}{4}\bar{g}_{ij}\bar{R}\right) &=& \mathcal{A}^2 \Big[
\delta_{ij} \left( \frac{\psi''}{2} + \frac{1}{2} (B')_{, l}^{\ l} +\frac{1}{2} A_{, l}^{\ \ l} - \frac{1}{4} (D_l^k E)^{, l}_{\ \ k} \right) 
-B'_{,ij} +\psi_{,ij} \nonumber \\ 
&-&  A_{,ij} +\frac{1}{2}D_{i j}E'' + \frac{1}{2}
(D^k_jE)_{,ik} + \frac{1}{2}
(D^k_i E)_{,jk} -\frac{1}{2}(D_{ij})^{,k}_{\ \ k}E\Big], \\
\mathcal{A}^2\left(\bar{R}_{i0}- \frac{1}{4}\bar{g}_{i0}\bar{R}\right)&=& \mathcal{A}^2 \left( 2  \psi '_{,i} + \frac{1}{2} ( D^k_i E')_{,k}\right).
\ea
The $(0,0)$, $(i,j)$ and $(i,0)$ components  of the perturbed part of other terms, $(\mathcal{A}^2)_{;\mu \nu} - \frac{1}{4} \bar{g}_{\mu\nu}(\mathcal{A}^2)_{;\lambda}^{ \ \ \lambda} $,
$\partial_\mu \mathcal{A}\partial_\nu {A}- \frac{1}{4} \bar{g}_{\mu\nu}\partial^\lambda \partial_\lambda \mathcal{A}$, and $T_{\mu\nu}
- \frac{1}{4} \bar{g}_{\mu\nu} T $, are given by 
\ba
(\mathcal{A}^2)_{;00} - \frac{1}{4} \bar{g}_{ 00}(\mathcal{A}^2)_{;\lambda}^{ \ \ \lambda} &=& -2 \mathcal{A} \mathcal{A}'  A' + 
\frac{ \mathcal{A}\mathcal{A}'}{2}\left( A' +  B_{,i}^{\ \ i} + 3 \psi' - \frac{1}{2} D^i_i E' \right) \nonumber \\
&=& \frac{ \mathcal{A}\mathcal{A}'}{2}\left(-3 A' +   B_{,i}^{\ \ i} + 3 \psi' - \frac{1}{2} D^i_i E' \right), \\
(\mathcal{A}^2)_{;ij} - \frac{1}{4} \bar{g}_{ ij}(\mathcal{A}^2)_{;\lambda}^{ \ \ \lambda} &=& 2  \mathcal{A}\mathcal{A}' 
\left( B_{,ij} + \delta_{ij}\psi'- \frac{1}{2}D_{ij}E'\right) + \frac{1}{2}(\mathcal{A}'^2 +\mathcal {A A}'')D_{ij}E \nonumber \\
&-& \delta_{ij} \Big[ (\mathcal{A}'^2 + \mathcal{ A A}'')(\psi + A)+ \frac{\mathcal{A A}'}{2} \left(
A'+ B_{,l}^{\ \ l} + 3 \psi' -\frac{1}{2} D^k_k E'\right)\Big], \nonumber \\
 \\
(\mathcal{A}^2)_{;i0} - \frac{1}{4} \bar{g}_{ i0}(\mathcal{A}^2)_{;\lambda}^{\ \ \lambda} &=& - 2 \mathcal{A A}'  A_{,i} +
\frac{1}{2} \left( \mathcal{A}'^2 + \mathcal{ A A}''\right) B_{,i} \ , \\
\partial_0 \mathcal{A}\partial_0 {A}- \frac{1}{4}\bar{g}_{00}\partial^\lambda \partial_\lambda \mathcal{A} &=&0 ,\\
\partial_i \mathcal{A}\partial_j {A}- \frac{1}{4}\bar{g}_{ij}\partial^\lambda \partial_\lambda \mathcal{A} &=&
-\frac{1}{2}\mathcal{A}'^2 \Big[ (\psi + A) \delta_{ij} - \frac{1}{2} D_{ij}E\big] ,\\
\partial_i \mathcal{A}\partial_0 {A}- \frac{1}{4}\bar{g}_{i0}\partial^\lambda \partial_\lambda \mathcal{A} &=& \frac{\mathcal{A}'^2}{4} B_{,i} \ ,\\
T_{00} - \frac{1}{4} \bar{g}_{00} T &=& \frac{3}{2} \mathcal{A}^2 \chi_0'\chi_1',\\
T_{ij} - \frac{1}{4} \bar{g}_{ij} T &=& \delta_{ij}  \frac{\mathcal{A}^2}{2} \left(-\chi_0'^2 \psi -\chi_0'^2 A + \chi_0' \chi_1'\right)
+ \frac{\mathcal{A}^2 \chi_0'^2}{4}D_{ij}E,\\
T_{i0} - \frac{1}{4} \bar{g}_{i0} T &=& \mathcal{A}^2 \left( \chi_0'  \chi_{1,i} +
\frac{1}{4} \chi_0'^2 B_{,i}\right). \label{0ipart4}
\ea
 Varying the matter action (\ref{matter_action}) with respect to the scalar field, we have 
 \ba
 \bar{D}^{\mu}\bar{D}_{\mu}\phi + 2 \frac{\partial^{\mu}{\mathcal{A}}}{A}\partial_{\mu} \phi - m^2 \mathcal{A}^2 \phi=0.
 \ea 
Using $\mathcal{A}= a(t)$, we obtain the  zeroth
and the first order equations as
\ba
\chi_0'' + \frac{2 a'}{a} \chi_0' + m^2 a^2 \chi_0 = 0 \label{scalar_lead},
\ea 
\ba
\chi_1''(k,\eta) + \frac{2 a'}{a}\chi_1'(k,\eta) - 2 A(\eta, k) \left(2 \frac{a'}{a}\chi_0' + \chi_0'' \right) + m^2 a^2 \chi_1(k,\eta) + k^2 \chi_1(k,\eta) = 0, \label{scalar_pert0}
\ea
where we expressed the perturbations in terms of the fourier mode such as 
\ba
\chi_1(\eta, \vec{x}) = \int \frac{d^3k}{(2\pi)^{3/2}} \chi_1(\eta, \vec{k}) e^{i \vec{k}.\vec{x}}.
\ea
 Using Eq. (\ref{scalar_lead}), we can simplify Eq. (\ref{scalar_pert0}) as
\ba
\chi_1'' + \frac{2 a'}{a}\chi_1' + k^2\chi_1 + m^2 a^2 \left(2 \chi_0 A + \chi_1\right) =0. \label{scalar_pert}
\ea
This equation is slightly different from the standard one because of the  constraint on $\sqrt{-g}$. 
The  constraint reduces the gauge degree of freedom. In our case, we cannot set both $E$ and $B$ to zero. We can set only one of these fields to zero
since we have the constraint Eq. (\ref{constraint1}). The comoving curvature is still a relevant quantity since it is independent of $E$ and $B$.
In the next section, we solve for the power spectrum of the scalar perturbation choosing a gauge $E=0 \ (B\neq0)$.

\section{Power Spectrum}\label{power}
 In this section, we solve Eq. (\ref{scalar_pert}) for the scalar perturbation. In the gauge $E=0$, we have $A=3 \psi$ from Eq. (\ref{constraint1}). Using
  Eqs.  (\ref{00part1})-(\ref{0ipart4}), we can write all the components of  Eq. (\ref{eq_Rmunu}).  The $(i,j) (i\neq j)$ and $(0,i)$ components of Eq. (\ref{eq_Rmunu}) are
  useful in eliminating the field $A$ in 
 Eq. (\ref{scalar_pert}) and can be written as
 \ba
 \psi- A = B' - 2\mathcal{H} B  ,
 \ea
\ba
\psi' - \mathcal{H} A + \frac{1}{4} \Big[\mathcal{H}^2 \left(1-\frac{\xi}{2}\right) + \frac{a''}{a} + 4 \pi G \chi_0'^2\Big] B = -  4\pi G \chi_0' \chi_1, \label{eqn0i}
\ea
where, $\mathcal{H}= a'/a$. Using Eqs. (\ref{eq3_A}) and (\ref{eqn0i}), we have
\ba
A \approx \frac{4 \pi G \chi_0'}{\mathcal{H}} \chi_1 \label{eqnA},
\ea
where we assume that the gravitational potential does change much ($\psi'\approx 0$) in the super-horizon scale. Plugging  $A$ in Eq. (\ref{eqnA}) into Eq. (\ref{scalar_pert}),
we obtain
\ba
\chi_1'' + 2 \frac{a''}{a}\chi_1' + k^2 \chi_1 + m^2a^2 \left(1 + s \right) \chi_1 \approx 0, \label{scalar_pert1}
\ea
where $s= 2 \chi_0 \sqrt{4 \pi G \epsilon_1}\ll 1$. Introducing $\sigma= a \chi_1$, we can rewrite   Eq. (\ref{scalar_pert1}) as
\ba
\sigma''+ \Big[k^2 - \frac{a''}{a} + \left(1+ s\right) m^2 a^2 \Big]\sigma \approx0. \label{scalar_pert2}
\ea
 In the quasi de-Sitter expansion ($\dot{H}= -\epsilon H^2$), $a(\eta)= - 1/ (1-\epsilon) H \eta$, one can have 
 \ba
 \frac{a''}{a}= a^2 \left(2 H^2 + \dot{H}\right) \approx \frac{2 + 3\epsilon}{\eta^2},
 \ea
 and,
 \ba
  - \frac{a''}{a} + \left(1+ s\right) m^2 a^2  &\approx& - \frac{2 + 3 \epsilon }{\eta^2}  +  \left(1+ s\right) (1 + 2 \epsilon)\frac{m^2}{H^2 \eta^2} \nonumber \\
  &\approx& -\frac{2 + 3 \epsilon - 3 \eta_{\chi}}{\eta^2}= -\left(\nu_{\chi}^2 -\frac{1}{4}\right)\frac{1}{\eta^2},
 \ea
 where $\eta_{\chi} = m^2/3 H^2 \ll 1$ and $\nu_{\chi} = 3/2 + \epsilon - \eta_{\chi}$. Therefore, Eq. (\ref{scalar_pert2}) can be expressed 
 in the following form, 
 \ba
 \sigma'' + \Big[k^2 - \left(\nu_{\chi}^2 - \frac{1}{4}\right)\frac{1}{\eta^2} \Big]\sigma \approx0.
 \ea
 The solution to this equation is given by,
 \ba
 \sigma= \sqrt{-\eta} \Big[c_1(k) H^{(1)}(-k\eta) + c_2(k)H^{(2)}(-k\eta)\Big],
 \ea
 where $H^{(1,2)}$ is the Hankel function of first and second kind. In the ultraviolet regime, $k\gg a H (-k\eta\gg 1)$, we have
 \ba
 H^{(1)} (-k \eta\gg 1)  \approx \sqrt{-\frac{2}{\pi k \eta}} e^{i\left( -k\eta -\pi\nu_{\chi}/2-\pi/4 \right)}, \ \ 
 H^{(2)} (-k \eta\gg 1)  \approx \sqrt{-\frac{2}{\pi k \eta}} e^{-i\left( -k \eta -\pi\nu_{\chi}/2-\pi/4 \right)}.
 \ea
  After imposing the boundary condition $c_2(k)=0$ and $c_1(k)= \left(\sqrt{\pi}/2\right) e^{i\left(\pi\nu_{\chi}/2 + \pi/4\right)}$, 
 the solution becomes a plane wave $e^{-ik\eta}/\sqrt{2 k}$. Then we have 
 \ba
 \chi_1\approx \frac{H}{\sqrt{2 k^3}} \left(\frac{k}{a H}\right)^{3/2 -\nu_{\chi}}, \label{solchi1}
 \ea
 and  the spectral index $n_{s}$ is given by
 \ba
 n_{s}= 1+ \frac{d \ln(P_{\chi})}{d \ln(k)}= 1+ 2 \eta_{\chi}- 2\epsilon .
 \ea
 The comoving curvature  $\mathcal{R}= \psi+  (\mathcal{H} \chi_1)/\chi_0'=A/3 +  (\mathcal{H}\chi_1)/\chi_0' \approx
( \mathcal{H}\chi_1)/\chi_0'$ is still a gauge invariant quantity since all the gauge transformations of perturbations which make the line element $ds_g$
 invariant, are  not related with $\sqrt{-g}$. The power spectrum  is given by
 \ba
P_{\mathcal{R}} = \frac{k^3}{2 \pi^2} \mathcal{R}^2 &\approx& \frac{k^3}{2 \pi^2} \frac{H^2 \chi_1^2}{\dot{\chi}_0^2}\approx \frac{k^3 H^2}{2 \pi^2 \dot{\chi}_0^2}\times \frac{H^2}{2 k^3}
\left(\frac{k}{a H}\right)^{3-2\nu_{\chi}} \nonumber \\
&\approx& \frac{H^2}{\dot{\chi_0}^2} \left(\frac{H}{2 \pi}\right)^2 \left(\frac{k}{a H}\right)^{3-2 \nu_{\chi}} \approx \frac{4 \pi G}{\epsilon_1} 
\left(\frac{H}{2 \pi}\right)^2 \left(\frac{k}{a H}\right)^{3-2 \nu_{\chi}}  \nonumber \\
&\approx&  \frac{4 \pi G}{\epsilon_1} \left(\frac{H}{2 \pi}\right)^2 .
 \ea
 Here, we note that the observed curvature perturbation is dependent of $\epsilon_1$ which is a function of the parameter $\delta$. The action corresponding to the tensor perturbation 
 is independent of the parameter $\xi$. Due to the traceless property of the tensor perturbation, $\sqrt{-g}$ is fixed up to the linear order, and thus
 it does not affect the equation of the  tensor perturbation up to the linear order.  Therefore, the tensor power spectrum is the same as in the standard one,
 \ba
 P_{\text{T}} = 64 \pi G \left(\frac{H}{2 \pi}\right)^2 \left(\frac{k}{a H}\right)^{3-2 \nu_T} \approx  64 \pi G \left(\frac{H}{2 \pi}\right)^2,
 \ea
 where $\nu_{\text{T}} \sim  3/2$.  The tensor-to-scalar ratio $r$ is then given by
 \ba
 r = \frac{P_{\text{T}}}{P_{\mathcal{R}}} \approx 16 \epsilon_1 \approx 32 \left(1+ \frac{ \delta}{6}\right)^{2}\left(1+ \frac{ \delta}{4}\right)^{2} \frac{ M_P^2}{\chi_{0}^2}  \approx \left(1+ \frac{ \delta}{6}\right)^{2}\left(1+ \frac{ \delta}{4}\right)^{2} r_{\mbox{\tiny GR}},
 \ea
 where $r_{\mbox{\tiny GR}}\sim 0.131$ is the tensor-to-scalar ratio in  GR  [$\xi=6$ $(\delta =0)$]. One can have $r<r_{\rm GR}$ when $-10\leq\delta \leq 0$.  Requiring $H^2>0$ and
 $\ddot{a}>0$ constrains the parameter further as $\delta> -2$ ($\xi>4$). In order to have a sufficient period of inflation,  we consider $|\delta|<1$.   From the recent 
 analyses of Planck Collaboration \cite{Adam:2015rua}, the value is expected as $r \leq 0.09$  which puts an upper bound on parameter as $\delta\lesssim -0.43$  ($ \xi \lesssim 5.57$).

\section{Conclusions} \label{conclusion}
In this paper, we have shown that classical picture of inflation can be reproduced by the unimodular theory of gravity, and have developed it in 
the perspective of the cosmological perturbation during inflation.  The unimodular theory is a subspace of the general theory of relativity. It reduces the number of the gauge degree of
freedom.  We have considered a  gauge condition $E=0 \ (B\neq 0)$ which is similar to the standard longitudinal gauge $E=B=0$. We obtained the similar 
perturbation relation as in the standard case. The tensor perturbation is not altered in this theory since the standard tensor perturbation 
fortunately satisfies the constraint $\sqrt{-g}$ fixed up to the linear order due to its traceless property, and also it is independent of the unimodular 
parameter $\xi$.  

We have shown that for $\xi=6$, all the physical quantities at the background level as well as at the perturbation level are the same as the standard 
GR. For $\xi\neq 6$, the number of $e$-foldings is modified at the same scale of inflaton field. The value
of the Hubble parameter increases as the $\xi$ decreases from 6.  The tensor-to-scalar ratio $r=16 \epsilon_1$ can
be lowered simultaneously. We found from the background equation that the physical accepted range is $4<\xi\leq6$. For the range
of the tensor-to-scalar ratio $r\leq 0.09$ from the new analyses of Planck collaboration \cite{Adam:2015rua},
we have found upper bound on $\xi$ is $ \sim 5.57$.

The enhanced gravitational wave has been observed  by BICEP2 collaboration \cite{Ade:2014xna}. 
Recently, Planck collaboration suggested the possibility of the enhanced signal is due to dust contaminations  \cite{Adam:2015rua},  but it still needs to be confirmed by new experiments.
Since we introduced an additional parameter $\xi$  in this work, fitting the values of the tensor-to-scalar ratio and the spectral index is more flexible. This value of $\xi$ will be further
constrained by forthcoming observations. Detailed study of the scalar and the tensor perturbation up to the second order will be pursued in the future publication.

\vspace{12pt}
\noindent {\bf Acknowledgements}

Authors are grateful to  M. Sami  for useful discussions  and to
the Centre for Theoretical Physics, JMI for the hospitality (N.K.S). This work was supported by the grant
from the National Research Foundation funded by the Korean government, No. NRF-2012R1A1A2006136.

\begin{spacing}{1}
\begin{small}

\end{small}
\end{spacing}
\end{document}